# Apparent spectral shift of thermally generated surface phonon-polariton resonance mediated by a non-resonant film

Vahid Hatamipour[*], Sheila Edalatpour[**] and Mathieu Francoeur[*,†]

[*]Radiative Energy Transfer Lab, Department of Mechanical Engineering, University of Utah, Salt Lake City, UT 84112, USA

[**]Department of Mechanical Engineering, University of Maine, Orono, ME 04469, USA

**ABSTRACT**

The physical origin of spectral shift of thermally generated surface phonon-polariton (SPhP) resonance of a silicon carbide (SiC) bulk mediated by a non-resonant film is elucidated. The local density of electromagnetic states (LDOS) in a non-resonant intrinsic silicon (Si) film due to thermal emission by SiC, derived using fluctuational electrodynamics, exhibits a local maximum near SPhP resonant frequency in addition to a lower frequency resonance generated by gap modes emerging in the vacuum gap separating the SiC and Si layers. Multiple reflections within the vacuum gap also induce a LDOS drop around SPhP resonant frequency. As a result, depending on the film thickness to vacuum gap ratio and the location where the LDOS is calculated in the film, the low-frequency resonance can dominate the LDOS, such that SPhP resonance appears to be redshifted. A similar spectral behavior is observed on the monochromatic radiative heat flux absorbed by the Si film. It is shown that apparent spectral (red and blue) shift of SPhP resonance mediated by a non-resonant film is bounded by the transverse and longitudinal optical phonon frequencies of SiC. This work is of importance in applications involving dissimilar materials, such as thermophotovoltaics and thermal rectification, where gap modes may significantly disrupt flux resonance. Gap modes may also be at the origin of the resonance redshift systematically observed in near-field thermal spectroscopy.

---

[†] Corresponding author. Tel.: + 1 801 581 5721
Email address: mfrancoeur@mech.utah.edu

# I. INTRODUCTION

Enhancement of radiation heat transfer in the near field beyond Planck's blackbody limit has been experimentally confirmed in various configurations, such as plane-plane [1-5], sphere-plane [6-8] and tip-plane [9]. The mechanisms responsible for this Super-Planckian radiative transfer are generally well understood. From a theoretical standpoint, near-field radiative transfer between similar resonant materials supporting surface polaritons in the infrared has been extensively studied [e.g., 10-14]. However, many potential applications of near-field thermal radiation such as thermophotovoltaics [15-19], thermal rectification [20-24], and flux modulation and amplification [25,26], often require dissimilar materials. While near-field radiative transfer between dissimilar resonant materials has been analyzed in the past [e.g., 20,26], the case of a resonant heat source interacting with a non-resonant layer has been essentially disregarded.

Near-field thermal spectroscopy, a promising method for characterizing near-field thermal spectra [27-34], also involves dissimilar materials. In this paradigm, a probing tip is brought within a sub-wavelength distance from a sample. By heating either the tip or the sample, the thermal near field between the tip and the sample is scattered to the far zone. The detected signal in the far zone is spectroscopically analyzed to extract the spectral distribution of energy density and local density of electromagnetic states (LDOS) of the sample. Spectral characterization of near-field thermal energy of materials supporting surface phonon-polaritons (SPhPs) in the infrared, namely silicon carbide (SiC), silicon dioxide and hexagonal boron nitride, have been performed using various probing tip materials (intrinsic silicon [27,29], tungsten [28], platinum-iridium [30]). In all these experiments, spectral redshifts of SPhP resonance of varying magnitudes (from ~ 5 cm$^{-1}$ to ~ 65 cm$^{-1}$) have been measured. In particular, O'Callahan et al. [29] experimented with a SiC sample and non-resonant probing tips made of intrinsic silicon (Si), and reported spectral redshifts varying from 5 cm$^{-1}$ to 50 cm$^{-1}$ with respect to SPhP resonance of SiC (948 cm$^{-1}$). Different models have been employed to interpret near-field thermal spectroscopy measurements [27,28,29,35,36]. However, there is a lack of study explaining the physical origin of resonance spectral shift arising when a non-resonant object is within the near field of a resonant heat source.

The objective of this work is therefore to study the impact of a non-resonant object on the spectral shift of surface polariton resonance. As the radiative flux transferred from a resonant heat source to a non-resonant object or the energy scattered in the far zone in near-field thermal spectroscopy is related to the LDOS within the non-resonant object, and not to the LDOS in vacuum, spectral distributions of LDOS generated within a non-resonant layer due to thermal emission by SiC supporting SPhPs in the infrared are analyzed. In order to elucidate the physics of near-field radiative transfer between resonant and non-resonant materials, a plane layer geometry for which an analytical solution for the LDOS can be derived is adopted. It is shown that apparent spectral blue and redshift of SPhP resonance arises due to gap modes in the vacuum gap separating the SiC and non-resonant layers. In addition, the impact of SPhP spectral shift is analyzed on the radiative flux absorbed by the non-resonant layer.

The rest of the paper is organized as follows. An expression for the LDOS within a non-resonant film made of a temporally non-dispersive and lossless material due to thermal emission by SiC is first derived. Next, spatial and spectral distributions of LDOS in the non-resonant layer are analyzed and interpreted via dispersion relations. Spectral distributions of radiative flux absorbed by the non-resonant film exhibiting apparent spectral shift of SPhP resonance are afterwards discussed. Concluding remarks are finally provided.

## II. LDOS IN A TEMPORALLY NON-DISPERSIVE AND LOSSLESS FILM

## A. Description of the framework

The impact of a non-resonant layer on the near-field thermal spectrum of SiC near SPhP resonance is analyzed using the schematic shown in Fig. 1. It is assumed that all interfaces are parallel, perfectly smooth, infinite along the $\rho$-direction and azimuthally symmetric with respect to $\theta$. The SiC and non-resonant layers, of thickness $t_1$ and $t_3$ respectively, are in local thermodynamic equilibrium, surrounded by vacuum and separated by a sub-wavelength gap of thickness $d$. Throughout this paper, SiC is emitting thermal radiation at a temperature $T_1 = 300$ K, while the non-resonant layer is assumed to be non-emitting ($T_3 = 0$ K). The LDOS is calculated within the non-resonant film at a distance $\Delta$ relative to the interface 2-3.

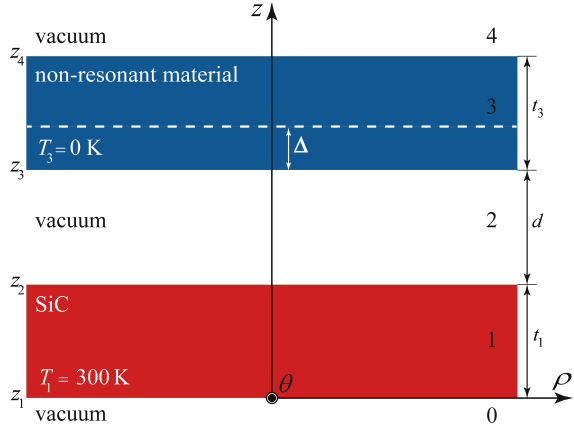

FIG. 1. Schematic of the problem under consideration. The LDOS due to an emitting SiC layer (medium 1, $T_1 = 300$ K) is calculated at a distance $\Delta$ relative to the interface 2-3 in the non-resonant film (medium 3, $T_3 = 0$ K). The SiC and non-resonant layers are surrounded by vacuum and are separated by a sub-wavelength gap of thickness $d$.

The spectral energy density, $u(\mathbf{r},\omega)$, in the non-resonant layer due to thermal emission by SiC can be expressed as the product of the spectral LDOS, $\rho(\mathbf{r},\omega)$, and the energy of an electromagnetic state $\Theta(\omega,T) = \hbar\omega/[\exp(\hbar\omega/k_B T)-1]$, where $\omega$ is the angular frequency, $k_B$ is the Boltzmann constant, and $\hbar$ is the reduced Planck constant [37,38]. The formulation is simplified hereafter by assuming that the non-resonant layer is non-magnetic, temporally non-dispersive and lossless. The time-averaged spectral energy density in the non-resonant layer can be written as [39]:

$$\langle u(\mathbf{r},\omega)\rangle = \frac{1}{4}\varepsilon_3'\varepsilon_0 \langle \mathbf{E}(\mathbf{r},\omega)\cdot\mathbf{E}^*(\mathbf{r},\omega)\rangle + \frac{1}{4}\mu_0 \langle \mathbf{H}(\mathbf{r},\omega)\cdot\mathbf{H}^*(\mathbf{r},\omega)\rangle \tag{1}$$

where $\mathbf{E}$ and $\mathbf{H}$ are the electric and magnetic field intensities, $\varepsilon_0$ and $\mu_0$ are the permittivity and permeability of vacuum, and $\varepsilon_3'$ is the real part of the dielectric function of the non-resonant film. A more general formulation for the electromagnetic energy density valid in temporally dispersive and lossy media can be found in Refs. [40-42]. The LDOS is derived using Eq. (1) and fluctuational electrodynamics [43]. The electric and magnetic fields are expressed in terms of a fluctuating current induced by thermal agitation, which is in turn correlated to the local temperature of the heat source, $T_1$, via the fluctuation-dissipation theorem. The details of the derivation are provided in the Appendix. Since this work is concerned by the near-field thermal spectrum of SiC near SPhP resonance, only TM-polarized evanescent modes in the non-resonant layer are considered. The TM evanescent component of the LDOS for the geometry shown in Fig. 1 is given by:

$$\rho_{\omega,13}^{evan,TM}(\Delta) = \frac{1}{2\pi^2\omega}\int_{\sqrt{\varepsilon_3'}k_0}^{\infty}\frac{k_{z2}''k_\rho^3\,dk_\rho}{|k_{z2}|^2}\frac{\left|t_{23}^{TM}\right|^2 \mathrm{Im}\left(R_1^{TM}\right)e^{2ik_{z3}\Delta}e^{2ik_{z2}d}}{\left|1+r_{23}^{TM}r_{34}^{TM}e^{2ik_{z3}t_3}\right|^2\left|1-R_1^{TM}R_3^{TM}\,e^{2ik_{z2}d}\right|^2}$$
$$\times\left[\left|1+r_{34}^{TM}e^{-2ik_{z3}(\Delta-t_3)}\right|^2 - 2\frac{k_{z3}''^2}{k_\rho^2}\mathrm{Re}\left(r_{34}^{TM}\right)e^{-2ik_{z3}(\Delta-t_3)}\right] \tag{2}$$

where $k_\rho$ and $k_{zj}$ are the parallel and perpendicular components of the wavevector with respect to the layer surfaces. The perpendicular wavevector in the non-resonant layer is calculated as $k_{z3} = \sqrt{\varepsilon_3' k_0^2 - k_\rho^2}$, where $k_0$ is the vacuum wavevector ($= \omega/c_0$). The integration over the parallel wavevector in Eq. (2) is performed for $k_\rho > \sqrt{\varepsilon_3'} k_0$, which implies that only modes that are evanescent in the non-resonant layer, characterized by purely imaginary perpendicular wavevectors $k_{z3} = ik_{z3}''$, are taken into account. The terms $r_{ij}^{TM}$ and $t_{ij}^{TM}$ are the Fresnel reflection and transmission coefficients at the interface between media $i$ and $j$, while $R_j^{TM} = \left( r_{j-1,j}^{TM} + r_{j,j+1}^{TM} e^{2ik_{zj}t_j} \right) / \left( 1 + r_{j-1,j}^{TM} r_{j,j+1}^{TM} e^{2ik_{zj}t_j} \right)$ is the reflection coefficient of layer $j$. Note that Eq. (2) can also be used for calculating the TE evanescent component of the spectral LDOS by replacing the superscripts TM by TE.

### B. Analysis of spectral LDOS in intrinsic Si

Spectral LDOS calculations are performed using a frequency-dependent dielectric function for SiC described by a Lorentz oscillator model $\varepsilon_1(\omega) = \varepsilon_\infty \left( \omega^2 - \omega_{LO}^2 + i\Gamma\omega \right) / \left( \omega^2 - \omega_{TO}^2 + i\Gamma\omega \right)$, where $\varepsilon_\infty = 6.7$, $\omega_{LO} = 1.825 \times 10^{14}$ rad/s, $\omega_{TO} = 1.494 \times 10^{14}$ rad/s, and $\Gamma = 8.966 \times 10^{11}$ 1/s [44]. In order to avoid complications associated with SPhP coupling within the heat source [13], it is assumed hereafter that the SiC emitter is optically thick such that it can be modeled as a semi-infinite bulk. Under this assumption, $t_1 \to \infty$ such that the reflection coefficient of layer 1, $R_1^{TM}$, reduces to the Fresnel reflection coefficient $r_{21}^{TM}$ in Eq. (2). SPhP resonance of the SiC bulk-vacuum interface can be determined using $r_{21}^{TM} \to \infty$ [39]. This condition is satisfied when $\varepsilon_1 k_{z2} + \varepsilon_2 k_{z1} = 0$. Solving this equation in the electrostatic limit where $k_\rho \gg k_0$, such that $k_{zj} \approx ik_\rho$, and neglecting losses in the dielectric function of SiC, SPhP resonance of a SiC-vacuum interface is given by $\omega_{res} \approx \sqrt{(\varepsilon_\infty \omega_{LO}^2 + \omega_{TO}^2)/(\varepsilon_\infty + 1)} = 1.786 \times 10^{14}$ rad/s (948 cm$^{-1}$). The case of a non-resonant film made of intrinsic Si is considered hereafter. This is justified by the fact that in the spectral band of interest ($1.6 \times 10^{14}$ rad/s to $1.9 \times 10^{14}$ rad/s) near SPhP resonance of a SiC-vacuum interface, the dielectric function of intrinsic Si shown in Fig. S1 of the Supplemental Material [45], $\varepsilon_3$ ($= \varepsilon_3' + i\varepsilon_3''$) $\approx 11.7 + i0.001$, is essentially frequency-independent and quasi-lossless ($\varepsilon_3'' \approx 0$) [44]. In addition, intrinsic Si is a material that has been used for probing LDOS in near-field thermal spectroscopy [27,29].

Spectral LDOS for Si film thickness to vacuum gap ratio, $D = t_3/d$, of 0.1, 1 and 10 are shown in Fig. 2 at different locations $\Delta/t_3$ within the Si layer for a vacuum thickness of $d = 10$ nm. SPhP resonant frequency of the SiC-vacuum interface is identified by a vertical dashed line. For $D = 0.1$, LDOS resonance is slightly redshifted, compared to SPhP resonant frequency of a SiC-vacuum interface, to a frequency of $1.782 \times 10^{14}$ rad/s (946.0 cm$^{-1}$) regardless of the location $\Delta/t_3$. When increasing $D$, a second low-frequency resonance emerges in addition to resonance near $\omega_{res}$. For both $D$ values of 1 and 10, the LDOS near SPhP resonant frequency significantly decreases as $\Delta/t_3$ increases, while the LDOS at the low-frequency resonance is a weak function of $\Delta/t_3$. As a result, at the back of the film (interface 3-4) where $\Delta/t_3$ is unity, SPhP resonance appears to be redshifted with respect to $\omega_{res}$ due to the dominance of the low-frequency mode. For $D = 1$ and 10, the low-frequency resonance occurs respectively at $1.754 \times 10^{14}$ rad/s (931.2 cm$^{-1}$) and $1.692 \times 10^{14}$ rad/s (898.2 cm$^{-1}$), which lead to apparent spectral redshifts of $0.032 \times 10^{14}$ rad/s (17 cm$^{-1}$) and $0.094 \times 10^{14}$ rad/s (49.9 cm$^{-1}$) with respect to SPhP resonance.

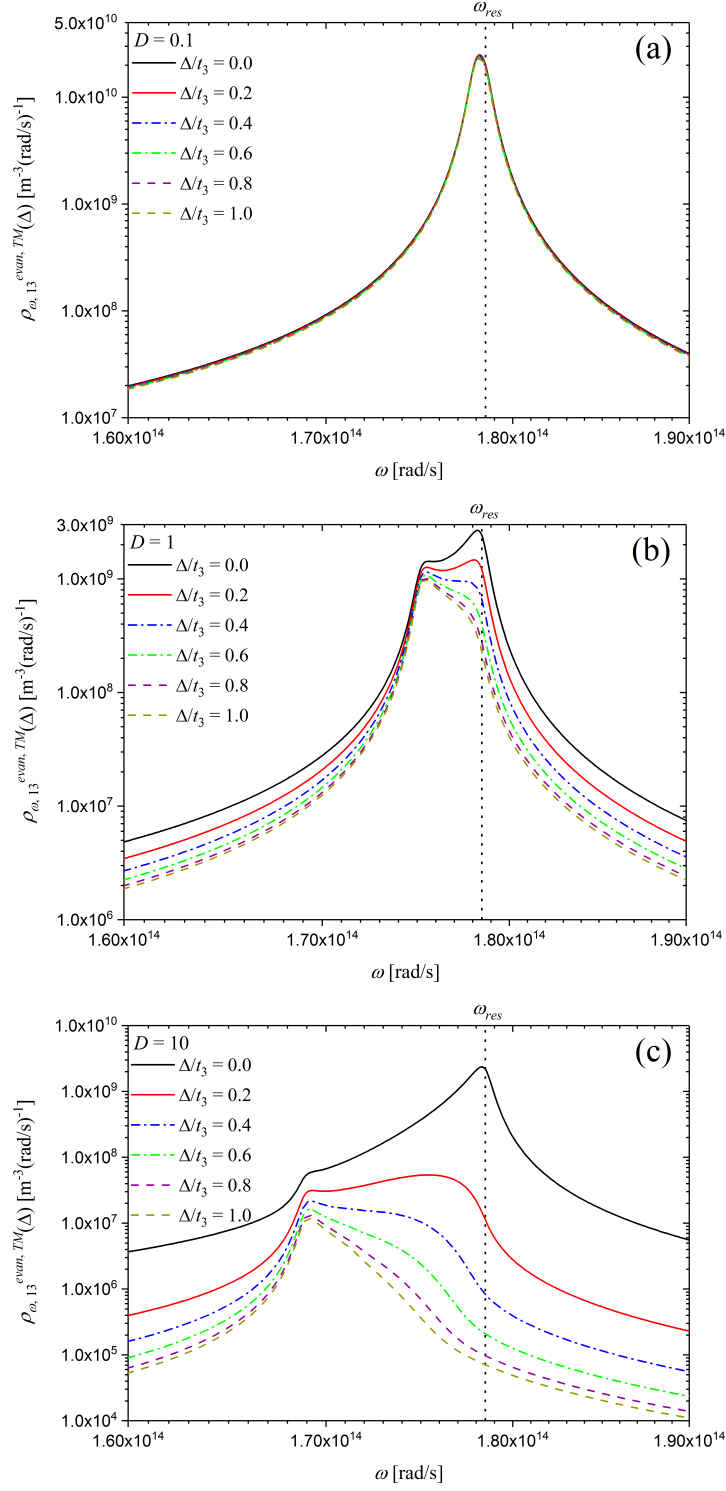

FIG. 2. TM evanescent component of the spectral LDOS as a function of $\Delta/t_3$ for $d = 10$ nm: (a) $D = 0.1$, (b) $D = 1$, and (c) $D = 10$. SPhP resonant frequency of a SiC-vacuum interface is identified in all panels by a vertical dashed line.

The physical origin of SPhP resonance apparent spectral redshift is investigated hereafter by calculating the dispersion relation for the bulk SiC-Si film configuration. Resonance of the spectral LDOS can occur either when $\operatorname{Im}(r_{21}^{TM}) \to \infty$ or when $\left|1 - r_{21}^{TM} R_3^{TM} e^{2ik_{z2}d}\right|^2 = 0$ (see Eq. (2), and apply the simplification $R_1^{TM} = r_{21}^{TM}$). As discussed above, the former condition provides the resonant frequency of a SiC-vacuum interface. The latter condition corresponds to gap modes in the vacuum gap of thickness $d$ separating the SiC and Si layers. The dispersion relation of the gap modes, $\omega_{gap}(k_\rho)$, is predicted in the electrostatic limit ($k_{zj} \approx ik_\rho$) and by neglecting losses in the dielectric function of SiC. The resulting dispersion relation is given by:

$$\omega_{gap}(k_\rho) \approx \sqrt{\frac{\left(\varepsilon_\infty \omega_{LO}^2 + \omega_{TO}^2\right)\left((\varepsilon_3'+1)^2 - (\varepsilon_3'-1)^2 e^{-2k_\rho t_3}\right) + \left(\omega_{TO}^2 - \varepsilon_\infty \omega_{LO}^2\right)\left((\varepsilon_3')^2 - 1\right)\left(1 - e^{-2k_\rho t_3}\right)e^{-2k_\rho d}}{(1+\varepsilon_\infty)\left((\varepsilon_3'+1)^2 - (\varepsilon_3'-1)^2 e^{-2k_\rho t_3}\right) + (1-\varepsilon_\infty)\left((\varepsilon_3')^2 - 1\right)\left(1 - e^{-2k_\rho t_3}\right)e^{-2k_\rho d}}} \qquad (3)$$

The terms leading to LDOS resonance, $\operatorname{Im}(r_{21}^{TM})$ and $\left|1 - r_{21}^{TM} R_3^{TM} e^{2ik_{z2}d}\right|^{-2}$, are plotted in Fig. 3 as a function of the frequency $\omega$ and the normalized parallel wavevector $K$ ($= k_\rho / \sqrt{\varepsilon_3'} k_0$). The term $\left|1 - r_{21}^{TM} R_3^{TM} e^{2ik_{z2}d}\right|^{-2}$, accounting for multiple reflections in the vacuum gap, is shown for $D$ values of 0.1, 1 and 10. SPhP resonance of a SiC-vacuum interface and the gap mode dispersion relation (Eq. (3)) are also included in the plots.

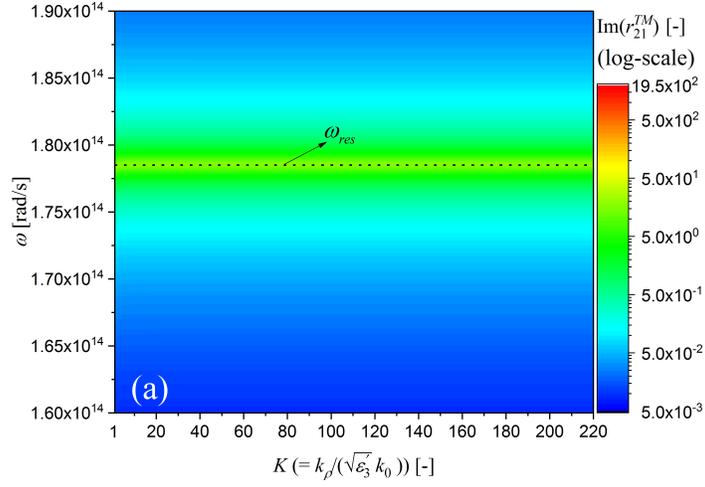

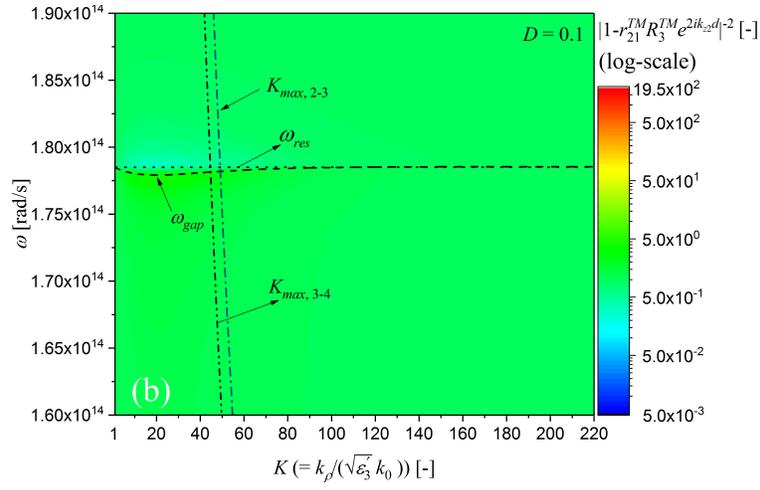

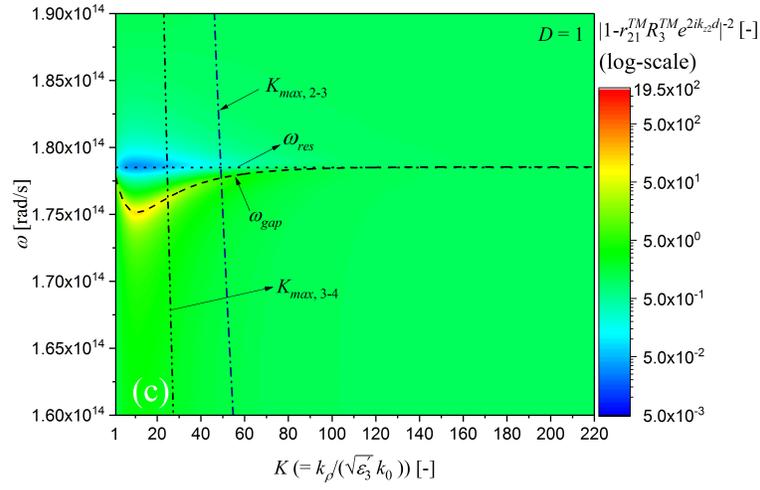

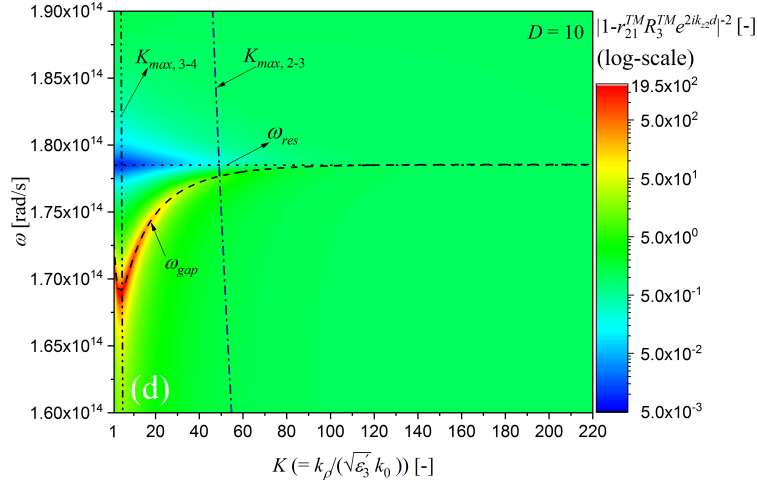

FIG. 3. Spectral distribution of $\text{Im}(r_{12}^{TM})$ (panel (a)) and $|1 - r_{12}^{TM} R_3^{TM} e^{2ik_{z2}d}|^{-2}$ (panels (b) to (d) for $D = 0.1$, 1 and 10) per unit parallel wavevector. In all cases, the vacuum gap thickness $d$ is 10 nm. For a better comparison, all contour plots have the same color range.

The term $\text{Im}(r_{21}^{TM})$ can be interpreted as the near-field emittance for evanescent waves in TM polarization of the SiC bulk, which is maximum at SPhP resonant frequency $\omega_{res}$ (see Fig. 3(a)). The multiple reflection term leads to a minimum around $\omega_{res}$ for small $K$ values. This minimum vanishes when $K$ is larger than approximately 40. In addition, the multiple reflection term induces a maximum matching a dip in the gap mode dispersion relation (see Figs. 3(b) to (d)). This maximum decreases in frequency and shifts toward smaller $K$ values as $D$ increases.

The effect of the gap modes on the spectral LDOS can be explained by considering the largest parallel wavevector contributing to the LDOS at a specific location in the Si film. At location $\Delta$ in the Si layer, only evanescent modes emitted by SiC with penetration depth equal to or larger than $(d + \Delta)$ can contribute to the LDOS. Using the definition of penetration depth, $\delta_j \approx |k_{zj}|^{-1}$, and the fact that $k_{zj} \approx ik_\rho$ in the electrostatic limit, the maximum contributing parallel wavevector is estimated as $k_{\rho,max} \approx (d+\Delta)^{-1}$ ($K_{max} \approx \left[\sqrt{\varepsilon'_3}k_0(d+\Delta)\right]^{-1}$). The maximum parallel wavevectors at the front and back surfaces of the Si film, $K_{max,2-3}$ and $K_{max,3-4}$, respectively, are plotted in Figs. 3(b) to (d). For $D = 0.1$, $K_{max,2-3}$ and $K_{max,3-4}$ are almost the same ($\approx 50$), thus leading to a spectral LDOS that is essentially invariant within the Si film (see Fig. 2(a)). At a $K$ value of approximately 50, the gap mode dispersion relation is slightly below $\omega_{res}$. This results in a small spectral redshift of SPhP resonance, as observed in Fig. 2(a).

For $D = 10$, the largest contributing parallel wavevectors to the LDOS at the front and back surfaces of the Si layer are approximately $K_{max,2-3} \approx 50$ and $K_{max,3-4} \approx 5$, respectively. For a $K$ value of 50, the minimum at $\omega_{res}$ induced by the multiple reflection term has almost vanished, while the maximum around the gap mode dispersion relation is fully contained within the $K$ interval from 1 to 50. This leads to a LDOS having a sharp resonance near $\omega_{res}$ and a smaller, low-frequency mode at $1.692 \times 10^{14}$ rad/s. At the back surface of the Si layer where $K_{max,3-4}$ is approximately 5, the situation is different. For $K$ values between 1 and 5, the minimum at $\omega_{res}$ induced by the multiple reflection term significantly reduces the LDOS at SPhP resonant frequency, even if the near-field emittance of the bulk SiC, $\text{Im}(r_{12}^{TM})$, is large. An important portion of the maximum around the gap mode dispersion relation is contained between 1 and 5, thus resulting in a strong, low-frequency resonance at $1.692 \times 10^{14}$ rad/s. The same interpretation is applicable to the case $D = 1$. As the thickness of the Si layer increases with respect to $d$, the low-frequency resonance further redshifts with respect to SPhP resonant frequency.

In summary, multiple reflections within the vacuum gap lead to an apparent spectral redshift of SPhP resonant frequency near the back surface of the film by inducing a drop of LDOS at SPhP resonant frequency as Δ increases, and by generating a low-frequency LDOS resonance that is a weak function of Δ. The Si film thickness to vacuum gap ratio, $D$, affects the gap mode dispersion relation, and thus affects the spectral location of the low-frequency mode. The impact of $D$ on the gap modes at the front and back surfaces of the Si film is shown in Fig. 4. In the limit that $D \to \infty$, the gap mode dispersion relation at the front and back surfaces of the film converges to frequencies of $1.777 \times 10^{14}$ rad/s (943.4 cm$^{-1}$) and $1.641 \times 10^{14}$ rad/s (871.2 cm$^{-1}$), respectively. Obviously, gap modes vanish in the limit that $D \to 0$ (i.e., absence of Si film).

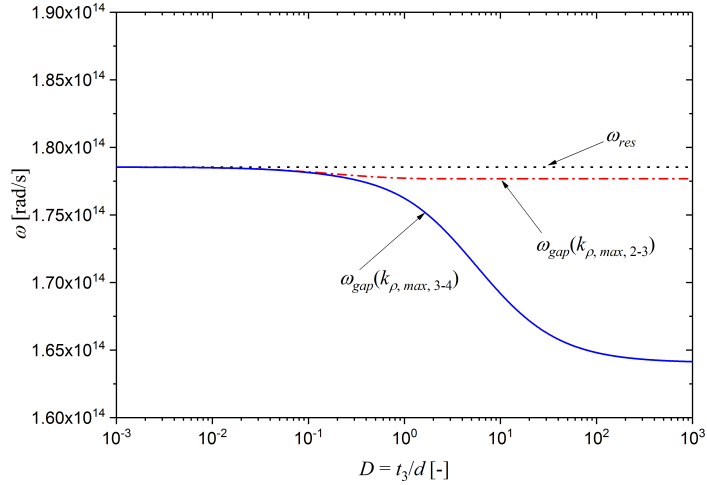

FIG. 4. Gap modes as a function of the film thickness to vacuum gap ratio, $D$, at the front ($\omega_{gap}(k_{\rho,max,2-3})$) and back ($\omega_{gap}(k_{\rho,max,3-4})$) surfaces of the intrinsic Si film. SPhP resonant frequency of a SiC-vacuum interface is identified by a horizontal dashed line.

### C. Limiting apparent spectral shifts of SPhP resonance

The conclusions made in the previous section can be generalized by varying the real part of the dielectric function of the non-resonant film, $\varepsilon'_3$. Using Eq. (3) in the hypothetical limit that $\varepsilon'_3 \to 0$, the gap mode dispersion relation is approximately equal to $\omega_{res}$ at the front surface of the film, independently of $D$; the same conclusion holds in the limit that $\varepsilon'_3 \to \infty$ (see Fig. S2(a) of the Supplemental Material [45]). At the back of the film, there is an apparent blueshift of SPhP resonance when $\varepsilon'_3$ takes values between 0 and 1. In the limit that $\varepsilon'_3 \to 0$ and $D \to \infty$, the gap mode dispersion relation converges to $\omega_{LO}$, which constitutes the limit of SPhP resonance apparent blueshift (see Fig. S2(b) of the Supplemental Material [45]). In a similar way, the limiting apparent redshift of SPhP resonance arises when $\varepsilon'_3 \to \infty$ and $D \to \infty$, where the gap mode dispersion relation converges to $\omega_{TO}$ (see Fig. S2(b) of the Supplemental Material [45]). The impact of small and large $\varepsilon'_3$ values (0.01 and 100) on LDOS profiles for a film thickness to vacuum gap ratio of unity is shown in Fig. S3 of the Supplemental Material [45].

Next, the impact of SPhP spectral shift on the monochromatic radiative heat flux absorbed by the non-resonant layer is discussed.

### III. NET RADIATIVE HEAT FLUX ABSORBED BY AN INTRINSIC SI FILM

The spatial and spectral distributions of LDOS in the Si film discussed in section II.B cannot be directly measured. Measurable quantities in near-field radiative heat transfer include scattered energy in the far zone and radiative flux. Here, the flat Si film cannot scatter in the far zone evanescent modes emitted by the SiC bulk. A grating or a micro/nanosize probe are required to couple evanescent modes into propagating modes in vacuum. As such, the overall impact of the spatial and spectral distributions of LDOS in the Si film on a measurable quantity is assessed hereafter by calculating the radiative flux. The TM evanescent component of the monochromatic radiative flux absorbed by the Si film due to thermal emission by the SiC bulk is given by [14,46,47]:

$$q_{\omega,13}^{evan,TM} = \frac{\Theta(\omega,T_1)}{\pi^2} \int_{\sqrt{\varepsilon_3'}k_0}^{\infty} k_\rho \, dk_\rho \, e^{2ik_{z2}d} \frac{\text{Im}(r_{21}^{TM}) \text{Im}(R_3^{TM})}{\left|1 - r_{21}^{TM} R_3^{TM} e^{2ik_{z2}d}\right|^2} \tag{4}$$

where only evanescent modes in Si are considered. Note that when calculating the flux, losses in the dielectric function of Si are taken into account using the data provided in Ref. [44] (see Fig. S1 of the Supplemental Material [45]).

Figure 5 shows the spectral distribution of radiative flux (TM evanescent component) for $D$ values of 0.1, 1 and 10. SPhP resonant frequency of the SiC-vacuum interface is identified by a vertical dashed line. For $D$ values of 0.1 and 10, SPhP resonance near $\omega_{res}$ dominates the radiative flux profiles. For $D = 1$, the flux around the low-frequency resonance mediated by gap modes at $1.754 \times 10^{14}$ rad/s (931.2 cm$^{-1}$) is larger than the flux near SPhP resonance. Thus, such a profile results in a large apparent SPhP resonance redshift of 17 cm$^{-1}$. This analysis therefore suggests that near-field thermal radiation applications capitalizing on dissimilar materials, such as thermophotovoltaics and thermal rectification, may be affected by gap modes inducing additional resonances in the flux. For a highly absorbing layer where the back surface of the film does not play any role, gap modes are expected to have a diminishing impact on the radiative flux. This is confirmed by Fig. S4 of the Supplemental Material [45], where the imaginary part of the dielectric function of the film, $\varepsilon_3''$, is increased up to 10 while keeping the real part, $\varepsilon_3'$, the same as for intrinsic Si. Apparent spectral redshift of SPhP resonance is visible on the flux up to a $\varepsilon_3''$ value of 1, and completely disappears for a large $\varepsilon_3''$ value of 10.

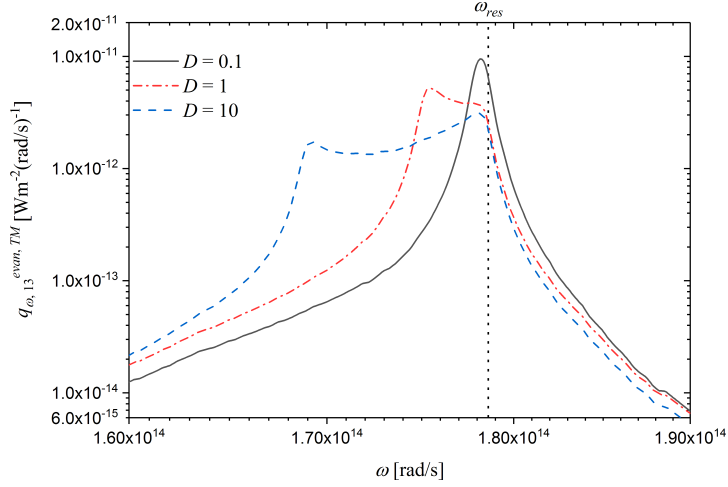

FIG. 5. TM evanescent component of the monochromatic radiative heat flux absorbed by medium 3 (Si film) due to thermal emission by medium 1 (SiC bulk, $T_1$ = 300 K) for $D$ = 0.1, 1 and 10. The vacuum gap thickness $d$ is fixed at 10 nm. SPhP resonant frequency of a SiC-vacuum interface is identified by a vertical dashed line.

## IV. CONCLUSIONS

The physical origin of surface polariton resonance spectral shift mediated by a non-resonant film was analyzed. For this purpose, a plane layer geometry where a non-resonant film located at a distance of 10 nm within the near-field thermal spectrum emitted by a SiC bulk was considered. For the case of intrinsic Si, the LDOS in the film exhibited a resonance at a frequency near SPhP resonance of a SiC-vacuum interface, in addition to a second, lower-frequency LDOS resonance emerging for film thickness to vacuum gap ratio $D$ of 1 and 10. The spectral location of the low-frequency LDOS resonance is a function of $D$, and its magnitude with respect to SPhP resonance depends on the location where LDOS is calculated in the Si film. Dispersion relation analysis revealed that the low-frequency resonance is generated by gap modes in the vacuum gap separating the SiC bulk and the Si film. In addition, multiple reflections within the vacuum gap induce a drop of LDOS near SPhP resonant frequency. These combined effects lead to an apparent spectral redshift of SPhP resonance as large as 49.9 cm$^{-1}$ for a $D$ value of 10. Similar conclusions were reached by calculating the monochromatic radiative flux absorbed by the Si film, where gap modes lead to low-frequency resonance and apparent spectral redshift of SPhP resonance. In general, for a polar crystal such as SiC, apparent spectral shift of surface polariton resonance mediated by a non-resonant film is bounded by the transverse and longitudinal optical phonon frequencies, $\omega_{TO}$ and $\omega_{LO}$. Resonance spectral shift is visible on flux profiles provided that losses in the non-resonant film are not too large (imaginary part of the dielectric function approximately equal to or smaller than 1).

The outcome of this work is important for applications capitalizing on dissimilar materials, such as thermal rectification and thermophotovoltaics, where gap modes may significantly affect resonance of the flux. In addition, despite the relative simplicity of the planar geometry, gap modes may explain the systematic resonance redshift measured in near-field thermal spectroscopy, since the scattered energy in the far zone is directly related to the LDOS in the probe. Accurate predictions of apparent spectral redshift in near-field thermal spectroscopy of the scattered energy in the far zone is beyond the scope of this work, and can potentially be obtained using numerical methods allowing precise modeling of the complex-shaped probing tip interacting with the sample [48].

## ACKNOWLEDGMENTS

This work was sponsored by the National Science Foundation under Grant No. CBET-1253577.

# APPENDIX: DERIVATION OF THE LOCAL DENSITY OF ELECTROMAGNETIC STATES (LDOS)

The LDOS is derived using fluctuational electrodynamics, where a fluctuating current $\mathbf{J}^{fl}$ representing thermal emission is added to Maxwell's equations [43]. The electric and magnetic fields generated at location $\mathbf{r}$ due to thermal emission by medium 1 (SiC) are given by:

$$\mathbf{E}(\mathbf{r},\omega) = i\omega\mu_0 \int_{V_1} \overline{\overline{\mathbf{G}}}^{E}(\mathbf{r},\mathbf{r}',\omega) \cdot \mathbf{J}^{fl}(\mathbf{r}',\omega) d^3\mathbf{r}' \tag{A1}$$

$$\mathbf{H}(\mathbf{r},\omega) = \int_{V_1} \overline{\overline{\mathbf{G}}}^{H}(\mathbf{r},\mathbf{r}',\omega) \cdot \mathbf{J}^{fl}(\mathbf{r}',\omega) d^3\mathbf{r}' \tag{A2}$$

where $\overline{\overline{\mathbf{G}}}^{E(H)}$ is the electric (magnetic) dyadic Green's function (DGF) relating the electric (magnetic) field observed at $\mathbf{r}$ to a source located at $\mathbf{r}'$. The ensemble average of the spatial correlation function of the fluctuating current is provided by the fluctuation-dissipation theorem [43]:

$$\left\langle J_\alpha^{fl}(\mathbf{r}',\omega) J_\beta^{fl*}(\mathbf{r}'',\omega') \right\rangle = \frac{4\omega\varepsilon_0 \varepsilon_1''(\omega)}{\pi} \Theta(\omega,T_1) \delta(\mathbf{r}'-\mathbf{r}'') \delta(\omega-\omega') \delta_{\alpha\beta} \tag{A3}$$

where $\alpha$ and $\beta$ specify the state of polarization of the fluctuating current.

The LDOS in medium 3 (assumed to be temporally non-dispersive and lossless) is derived by substituting the electric and magnetic fields into the spectral energy density (Eq. (1)) and by applying the fluctuation-dissipation theorem. After some manipulations, the spectral LDOS at location $\Delta$ in medium 3 is written as:

$$\rho_{\omega,13}(\Delta) = \frac{\omega \varepsilon_1''(\omega)}{2c_0^2 \pi^2} \int_0^{\infty} k_\rho \, dk_\rho \int_0^{t_1} \left( k_3^2 \left| g_{13m\alpha}^E (k_\rho, \Delta, z', \omega) \right|^2 + \left| g_{13m\alpha}^H (k_\rho, \Delta, z', \omega) \right|^2 \right) dz' \tag{A4}$$

where the subscript $m\alpha$ implies that a summation is performed over the components $\rho$, $\theta$, and $z$, while $g_{13m\alpha}^{E(H)}$ is the Weyl component of the electric (magnetic) DGF. The Weyl components of the electric DGF relating the fields originating from $z'$ in medium 1 to the fields observed at $\Delta$ in medium 3 are [49]:

$$\overline{\overline{\mathbf{g}}}_{13}^E(k_\rho, \Delta, z', \omega) = \frac{i}{2k_{z1}} \left[ \begin{array}{l} \left(A_3^{TE} \hat{\mathbf{s}}\hat{\mathbf{s}} + A_3^{TM} \hat{\mathbf{p}}_3^+ \hat{\mathbf{p}}_1^+\right) e^{i[k_{z3}\Delta - k_{z1}z']} + \left(B_3^{TE} \hat{\mathbf{s}}\hat{\mathbf{s}} + B_3^{TM} \hat{\mathbf{p}}_3^- \hat{\mathbf{p}}_1^+\right) e^{i[-k_{z3}\Delta - k_{z1}z']} \\ + \left(C_3^{TE} \hat{\mathbf{s}}\hat{\mathbf{s}} + C_3^{TM} \hat{\mathbf{p}}_3^+ \hat{\mathbf{p}}_1^-\right) e^{i[k_{z3}\Delta + k_{z1}z']} + \left(D_3^{TE} \hat{\mathbf{s}}\hat{\mathbf{s}} + D_3^{TM} \hat{\mathbf{p}}_3^- \hat{\mathbf{p}}_1^-\right) e^{i[-k_{z3}\Delta + k_{z1}z']} \end{array} \right] \tag{A5}$$

where $\hat{\mathbf{s}}$ and $\hat{\mathbf{p}}_j^\pm$ are TE- and TM-polarized unit vectors, respectively [50]. The coefficients $A_3^\gamma$ and $B_3^\gamma$ are amplitudes of forward ($z$-positive) and backward ($z$-negative) traveling waves in polarization state $\gamma$ in layer 3 generated by a source emitting in the forward direction. The same definition holds for the coefficients $C_3^\gamma$ and $D_3^\gamma$, except that the source is emitting in the backward direction. The Weyl components of the magnetic DGF are readily obtained from Eq. (A5) using $\overline{\overline{\mathbf{g}}}_{13}^H = \nabla \times \overline{\overline{\mathbf{g}}}_{13}^E$.

The amplitude coefficients needed to calculate the Weyl components of the electric and magnetic DGFs are determined using a transfer matrix approach [51]:

$$A_3^\gamma = \frac{t_{12}^\gamma t_{23}^\gamma e^{ik_{z1}t_1} e^{ik_{z2}t_2}}{\left(1 + r_{01}^\gamma r_{12}^\gamma e^{2ik_{z1}t_1}\right)\left(1 + r_{23}^\gamma r_{34}^\gamma e^{2ik_{z3}t_3}\right)\left(1 - R_1^\gamma R_3^\gamma e^{2ik_{z2}t_2}\right)} \tag{A6}$$

$$B_3^\gamma = r_{34}^\gamma e^{2ik_{z3}t_3} A_3^\gamma \tag{A7}$$

$$C_3^\gamma = -r_{01}^\gamma A_3^\gamma \tag{A8}$$

$$D_3^\gamma = r_{34}^\gamma e^{2ik_{z3}t_3} C_3^\gamma \tag{A9}$$

The spectral LDOS in medium 3 is obtained by including the amplitude coefficients in the Weyl components of the electric and magnetic DGFs, which are in turn substituted in Eq. (A4). The resulting expression is given by Eq. (2), where only the TM evanescent component in medium 3 has been kept.

Supplemental Material for article "Apparent spectral shift of thermally generated surface phonon-polariton resonance mediated by a non-resonant film"

Vahid Hatamipour[*], Sheila Edalatpour[**] and Mathieu Francoeur[*,†]

[*]*Radiative Energy Transfer Lab, Department of Mechanical Engineering, University of Utah, Salt Lake City, UT 84112, USA*

[**]*Department of Mechanical Engineering, University of Maine, Orono, ME 04469, USA*

(Dated: August 20, 2018)

[†] Corresponding author. Tel.: + 1 801 581 5721
Email address: mfrancoeur@mech.utah.edu

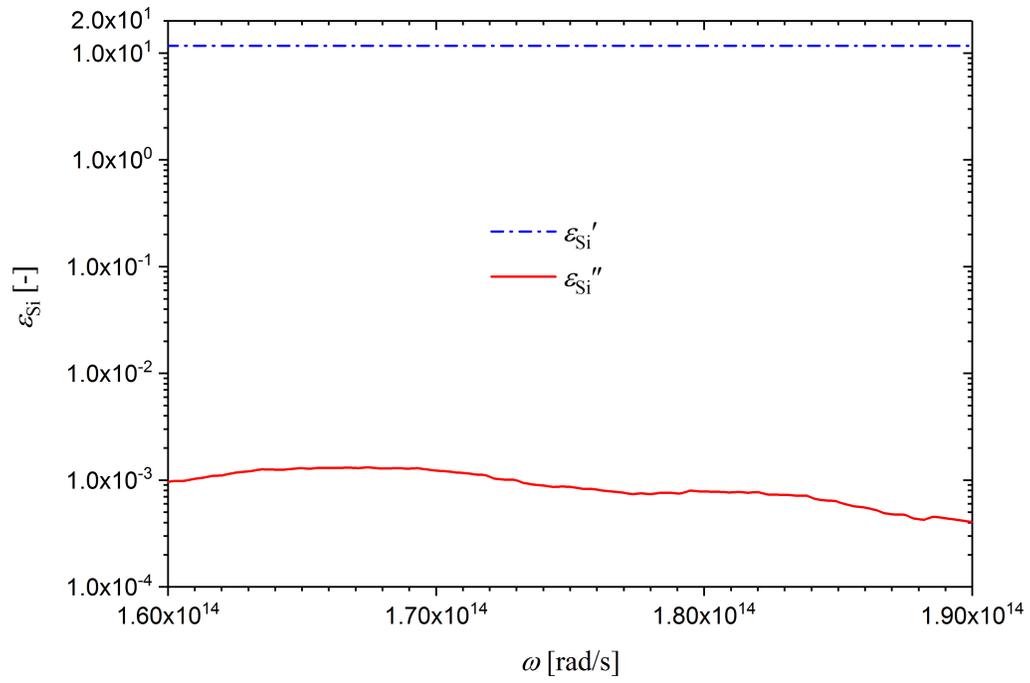

FIG. S1. Real ($\varepsilon'_{Si}$) and imaginary ($\varepsilon''_{Si}$) parts of the dielectric function of intrinsic Si.

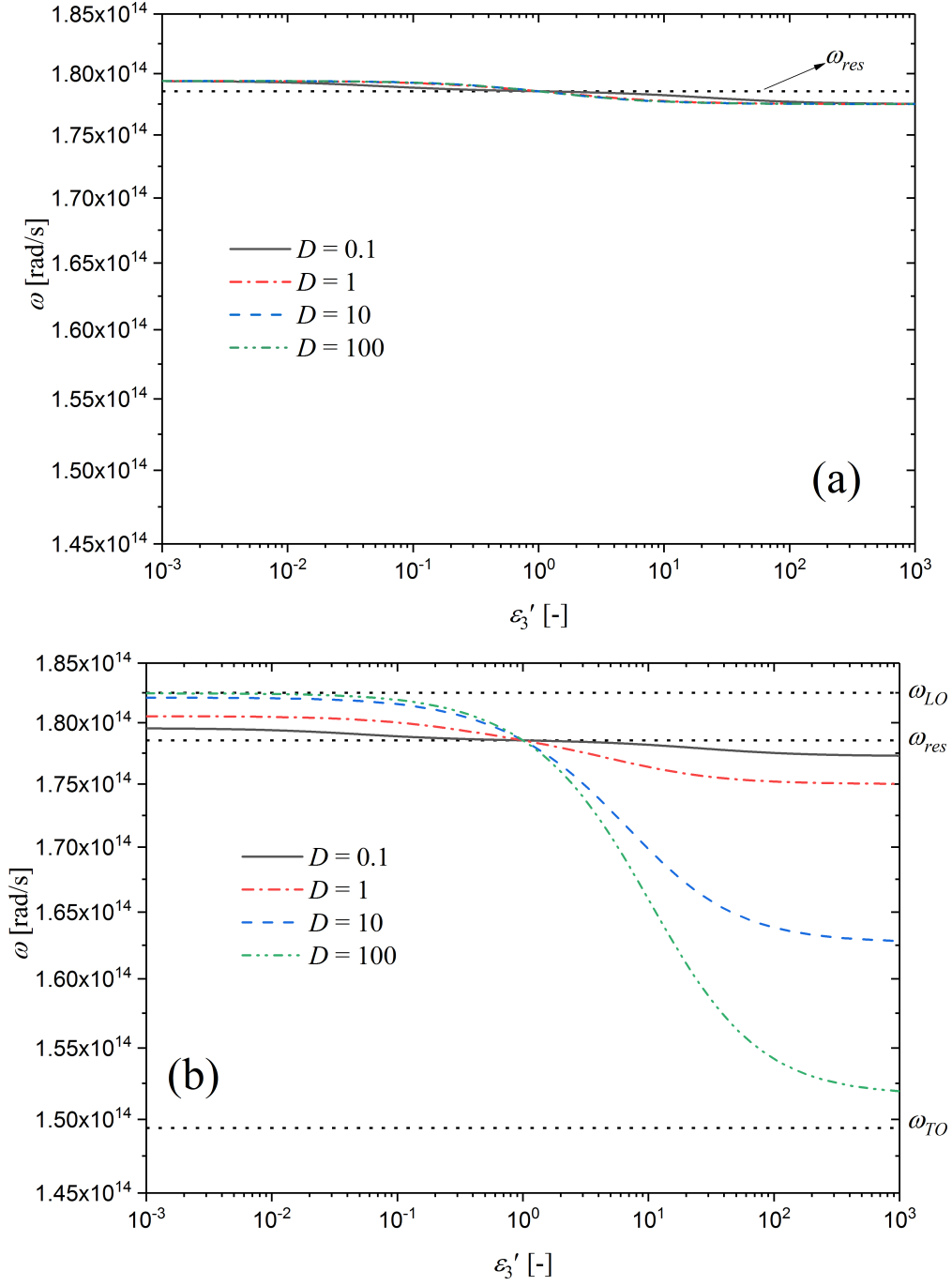

FIG. S2. Gap modes as a function of the real part of the dielectric function of the non-resonant layer, $\varepsilon'_3$, and the film thickness to vacuum gap ratio $D$ at the: (a) Front surface of the non-resonant layer, and (b) Back surface of the non-resonant layer.

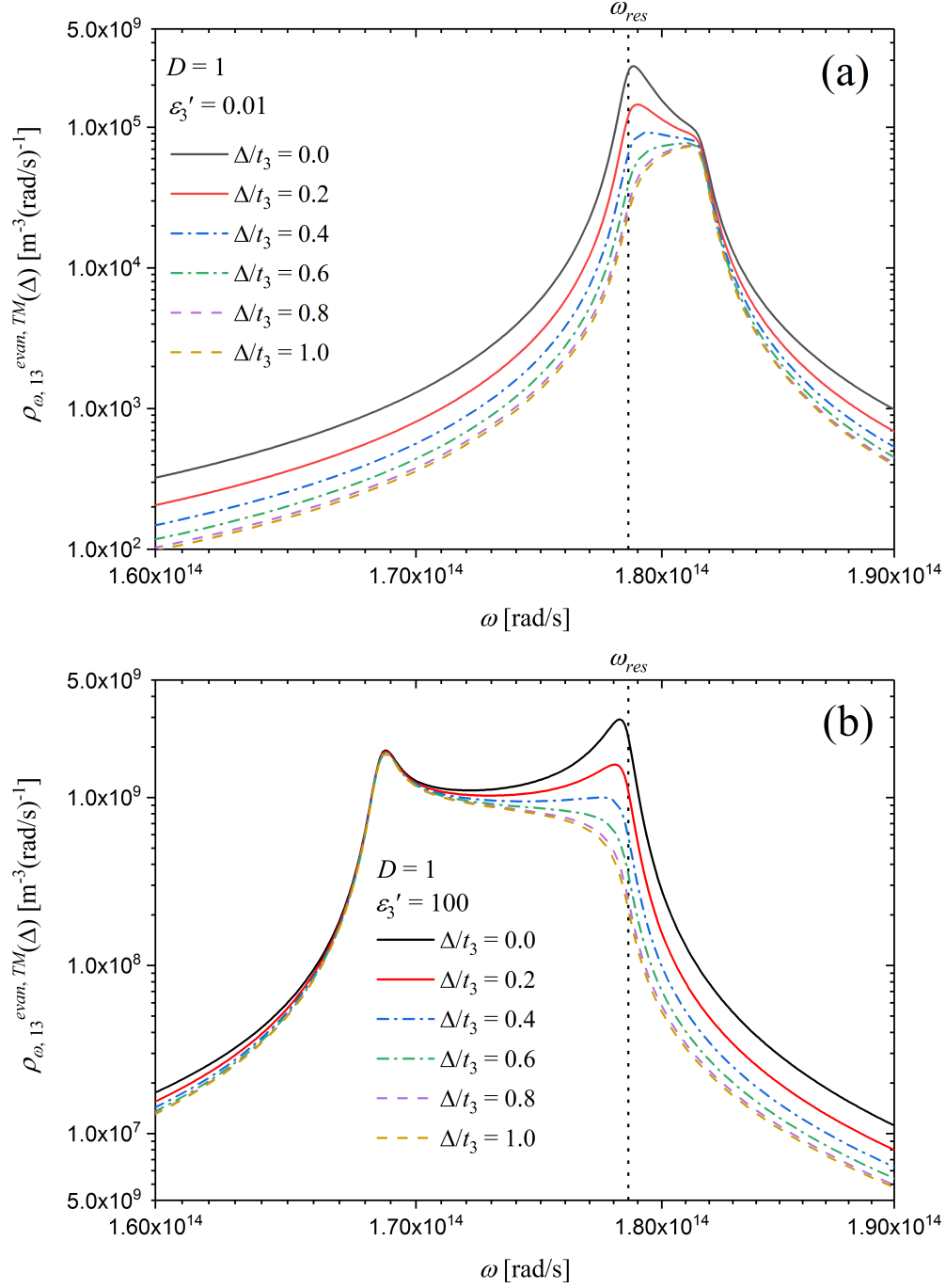

FIG. S3. TM evanescent component of the spectral LDOS as a function of $\Delta/t_3$ for $d = 10$ nm and $D = 1$: (a) $\varepsilon_3' = 0.01$, and (b) $\varepsilon_3' = 100$. SPhP resonant frequency of a SiC-vacuum interface is identified in all panels by a vertical dashed line.

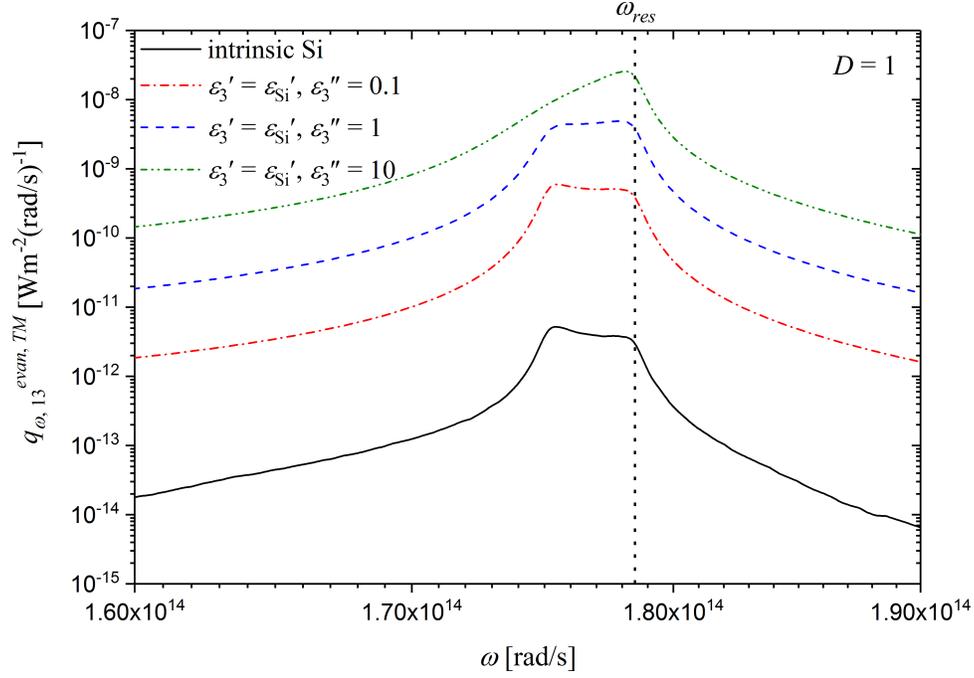

FIG. S4. TM evanescent component of the monochromatic radiative heat flux absorbed by medium 3 (non-resonant layer) due to thermal emission by medium 1 (SiC bulk, $T_1 = 300$ K) for $D = 1$. The vacuum gap thickness $d$ is fixed at 10 nm. The impact of losses is analyzed by increasing the imaginary part of the dielectric function of the non-resonant layer, $\varepsilon_3''$, while keeping the real part, $\varepsilon_3'$, the same as for intrinsic Si. SPhP resonant frequency of a SiC-vacuum interface is identified by a vertical dashed line.